\documentclass{aa}  

\usepackage{graphicx}
\usepackage{txfonts}
\begin{document}

   \title{The formation of urea in space}

   \subtitle{I. Ion-molecule, neutral-neutral, and radical gas-phase reactions}

   \author{Flavio Siro Brigiano
          \inst{1}
          \and
          Yannick Jeanvoine
          \inst{1}
          \and
          Antonio Largo
          \inst{2}
          \and
          Riccardo Spezia\inst{1}\fnmsep\thanks{present address: Sorbonne Universit\'es, UPMC Univ. Paris 06, UMR - CNRS 7616, Laboratoire de Chimie Th\'eorique, 75252 Paris, France.
          E-mail: riccardo.spezia@upmc.fr.}
          }

   \institute{LAMBE, Univ Evry, CNRS, CEA, Universit\'e Paris-Saclay, 91025, Evry, France\\
              \email{riccardo.spezia@univ-evry.fr}
         \and
        Computational Chemistry Group, Departamento de Quimica Fisica, Facultad de Ciencias, Universidad de Valladolid, Valladolid, Spain\\
             }

   \date{Received XX; accepted YY}

 
  \abstract
   {Many organic molecules have been observed in the interstellar medium thanks to advances in radioastronomy, and 
   very recently the presence of urea was also suggested. While those molecules were observed, it is not clear what the
   mechanisms responsible to their formation are. In fact, if gas-phase reactions are responsible, they should occur through barrierless mechanisms (or with very low barriers). In the past, mechanisms for the formation 
   of different organic molecules were studied, providing only in a few cases energetic conditions favorable to a synthesis at very low temperature. A particularly intriguing class of such molecules are those
   containing one N--C--O peptide bond, which could be a building block for the formation of biological molecules. Urea is a particular case because two nitrogen atoms are linked to the C--O moiety. 
   Thus, motivated also by the recent tentative observation of urea, we have
   considered the synthetic pathways responsible to its formation.}
   {We have studied the possibility of forming urea in the gas phase via different kinds of bi-molecular reactions: ion-molecule, neutral, and radical. In particular we have focused on the activation energy of these
   reactions in order to find possible reactants that could be responsible for to barrierless (or very low energy) pathways.}
   {We have used  very accurate, highly correlated quantum chemistry calculations to locate and characterize the reaction pathways in terms of  minima and transition states connecting reactants to products.}
   {Most of the reactions considered have an activation energy that is too high; but the ion-molecule reaction between NH$_2$OH$_2^+$ and formamide is not too high. These reactants could be responsible not only for the formation of urea
   but also of isocyanic acid, which is an organic molecule also observed in the interstellar medium.} 
   {}

   \keywords{ISM: molecules --
                astrochemistry --
                astrobiology --
                prebiotic molecules --
                reaction mechanisms
               }

   \maketitle
%

\section{Introduction}

In recent years a growing number of organic molecules have been detected in space, in particular in the interstellar medium (ISM) thanks to advances in radioastronomy, and in comets  (e.g.,
the recent mission on the P67/Churyumov-Gerasimenko comet already reported the presence of some interesting molecules like methyl isocyanate, acetone, acetamide, 
propionaldehyde, and glycine (\cite{Goesmann2015,Altwegge1600285}) and more are expected in the near future). 
A particularly interesting class of organic molecules is that
containing N--C single bonds, since these molecules can be the precursors for direct formation of peptides without passing from the single amino acids, as was initially suggested by~\cite{formamide} together with 
formamide (H$_2$NCHO) observation in the ISM. After formamide, other molecules containing  N--C bonds were found in the ISM, like 
cyanamide (H$_2$NCN,~\cite{Turner1975}), H$_3$CNC  (\cite{Cernicharo1988,Gratier2013}), 
acetamide (H$_2$NCOCH$_3$,~\cite{Hollis2006}), cyanoformaldehyde (CNCHO,~\cite{Remijan2008}), H$_2$NCO$^+$ (\cite{Gupta2013}), methylamine (H$_3$CNH$_2$,~\cite{CH3NH2,Fourikis1974}), 
aminoacetonitrile (H$_2$NCH$_2$CN,~\cite{Belloche2008}), and, very recently, urea ((H$_2$N)$_2$CO,~\cite{urea}) and N-methylformamide (H$_3$CNHCHO,~\cite{Belloche2017}). 
Urea  has the peculiarity of containing two N--C bonds and it is considered to be very important for the formation of life, since it can be a precursor for the synthesis of cytosine and uracil (\cite{kolb2014}). 
A tentative detection of urea in Sgr B2 molecular cloud was reported by collecting data from four independent astronomical instrumentations (\cite{urea}). 
Extraterrestrial urea was observed previously in the Murchinson C2 chondrite (\cite{hayatsu1975}) and
its presence was suggested  on grain mantles (\cite{raunier2004}). 

From the point of view of astrochemistry, the question that arises, related to the observation of these organic molecules, concerns the mechanisms that can be responsible for their synthesis. 
In general, low temperature environments promote only barrierless reaction pathways. Following this hypothesis, quantum chemistry calculations reported that while some possible mechanisms for the formation of formamide (\cite{Redondo2014}) and
glycine (\cite{Barrientos2012}) should be discarded due to high reaction barriers, others
are possible, like the formation of acetamide  through H$_2$NCHO + CH$_5^+$ reaction (\cite{Redondo2014b}) or formamide through a radical mechanism (\cite{Barone2015}). 
Ion-molecule collisions were suggested as a possible mechanism in both experimental (\cite{geppert2013,Larsson2012,blago2003,petrie2007,bohme}) and theoretical studies (\cite{Spezia2016}). In this case even reactions with (small) activation barriers are possible thanks to the 
transformation in the collision of some of the translational energy
into internal energy. Also gas-phase neutral-neutral reactions are a possible source of synthesis of carbon-bearing molecules (\cite{Kaiser2012}).

Another suggested way in which the synthesis of these molecules s that they were formed on interstellar grains surface. One particularly studied case is ice, which can decrease the activation energy.
For example,  the formation of glycine (\cite{Holtom2005}), amino acids (\cite{Elsila2007}), dipeptides (\cite{Keiser2013}), hydroxyacetonitrile, aminomethanol  and polyoxymethylene (\cite{Danger2012,Danger2014}),
glycolaldehyde and glyceraldehyde (\cite{deMarcellus2015}), or
alpha-aminoethanol (\cite{Duvernay2010}) in ice was reported experimentally. Recently, \cite{Forstel2016} have synthesized urea in cometary model ices thanks to the irradiation of inorganic ice (which contains CO and NH$_3$), 
proposing a mechanism in which
NH$_2\cdot$  is added to hydrogenated CO.

In many cases, calculations have shown that the activation energies are reduced by the presence of water molecules  (\cite{Woon99,Woon2001,Woon2002,Koch2008,Chen2011,Rimola2012}). 
Even if the activation barriers were reduced, they are still generally too high to allow a thermal reaction at low temperature conditions; however, in some cases they are compatible with 
the observed abundances in low-mass protostars and dark cores (\cite{Rimola2014}). 

In the present work, we have studied,  by highly correlated  quantum chemistry calculations, the possibility of forming urea by three classes of gas-phase reactions: ion-molecule, neutral-neutral, and radical. 
In the case of ion-molecule reaction, one tempting reactant to check is neutral and protonated hydroxylamine (NH$_2$OH). 
In particular, the protonated form was suggested to be a possible precursor for the formation
of glycine and formamide by ion-molecule collisions (\cite{bohme,Spezia2016}), but the potential energy surfaces always have an activation barrier (\cite{Barrientos2012,Redondo2014}). 
Despite the neutral form not being observed in the ISM, its protonated species could be formed and just not detected, as suggested by~\cite{Pulliam2012},  who provided
a clear picture of the non-detection of the neutral species suggesting the possibility of the existence of the protonated one. 
Our calculations have found  that the NH$_2$OH$_2^+$ + H$_2$NCHO (formamide) 
reaction  proceeds via a barrierless pathway, such that this is a plausible route for its synthesis under ISM conditions. 
We should note that NH$_2$OH$_2^+$ is the high energy form of protonated hydroxylamine:  calculations by~\cite{Boulet} have shown that protonation of neutral NH$_2$OH could provide both isomers (NH$_3$OH$^+$ 
and NH$_2$OH$_2^+$), 
whose  interconversion barrier is too high to be passed, so that the more reactive NH$_2$OH$_2^+$ species can exist.  
Other reactions involving species observed in the ISM are also investigated but reactions proceed through non-negligible energy barriers. 
Finally we have studied the neutral-neutral reaction and the radical-molecule reaction. Both have reaction barriers that are too high.

The remainder of the article is as follows.
The methods used are described in Section~\ref{sec:methods}, results are then presented first in terms of the thermodynamics of the reactions investigated (Section~\ref{sec:thermo}) and then we detail the
reaction pathways in terms of minima and transition states connecting reactants to products (Sections~\ref{sec:mech},~\ref{sec:mechneutral}, and~\ref{sec:radical}). Section~\ref{sec:concl} concludes.

\section{Computational details}\label{sec:methods}
  All structures (minima and transition states, TSs) were first optimized at the B3LYP/6-311++G(d,p) level of theory (\cite{Becke93,Lee88}), then, using these structures as starting points, we
located minima and TSs using the M{\o}ller-Plesset level of theory (MP2)  with aug-cc-pVTZ basis set (\cite{Dunning89,Davidson96}). 
   Finally, the electronic energies were calculated using the  highly correlated coupled cluster method, CCSD(T) (\cite{Purvis82,Pople87}),  with the same basis set. Energies are thus reported
   using the CCSD(T)/aug-cc-pVTZ electronic energy on the MP2/aug-cc-pVTZ minima and TSs. Zero point energy (ZPE) is calculated at the same MP2 level of theory. These results are thus 
   labeled as CCSD(T)/aug-cc-pVTZ//MP2/aug-cc-pVTZ in the following. 
   With structures of reactants and products, we calculated the energy differences taking into account the ZPE corrections and the Gibbs free energy differences at 15~K using standard
   thermochemistry  and the harmonic frequencies (\cite{McQuarrie73}).
   
   To verify that the TSs correctly connect the minima identified in the potential energy surface (PES), we
performed intrinsic reaction coordinate (IRC) calculations (\cite{Fukui81}). 
   All results show that the TSs correspond to the saddle point in the minimum energy path between two given minima (reactants, products and  intermediates). 
   
   In the case of radical reactions, we considered the doublet spin multiplicity and verified that there is no spin contamination. We obtained $\langle S\rangle^2 $ values very close to the
   theoretical one (0.75) and, for two transition states that have values about 10\% higher than the theoretical value, T1 diagnostics (\cite{Lee89})  on CCSD(T) calculations were performed,
   providing very small values, thus strengthening our confidence in  the calculation results. 
   
   Spontaneous emission rate constants were calculated using frequencies and Einstein  coefficients obtained by ab initio calculations and using the method reported by~\cite{Herbst82} 
   with our in-house code.
All quantum chemistry and thermochemistry calculations were performed using Gaussian 09 software (\cite{g09}).

\section{Results}
\subsection{Thermodynamics}\label{sec:thermo}

\begin{figure*}
\begin{center}
\includegraphics[width=1.0\textwidth]{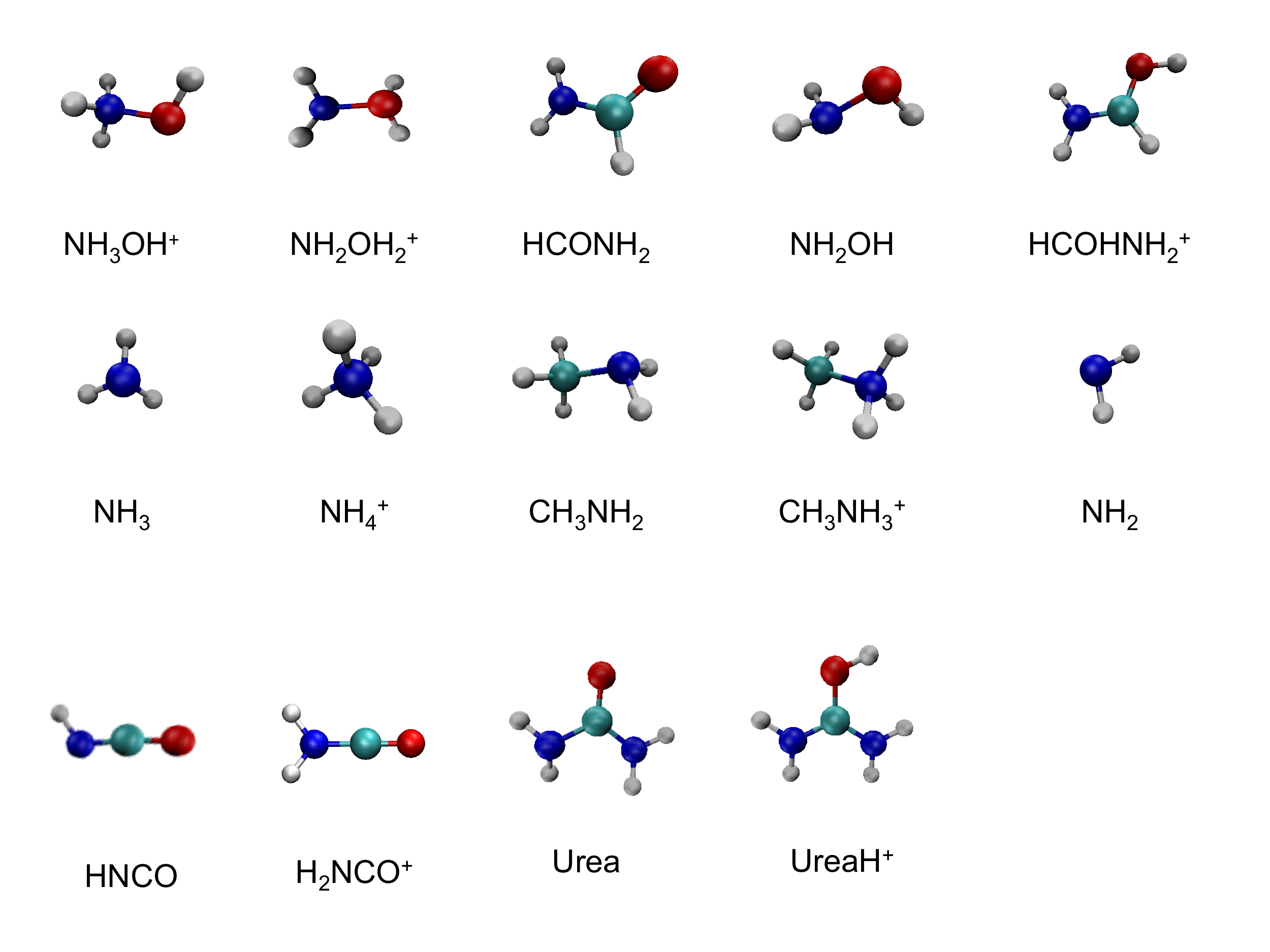}\\
\end{center}\caption[] {Structures of reactants and products considered (simple products like H$\cdot$, H$_2$, H$_2$O, and CH$_4$ are omitted). Oxygen atoms are in red, nitrogen in blue, carbon in cyan, and hydrogen in white.}
\label{fig:react-prod}
\end{figure*}

We first considered the energetics in terms of $\Delta E$ including the ZPE effects and gas phase $\Delta G$ at 15~K of possible reactions leading to urea. In particular we took into account three classes
of reactions, ion-molecule, neutral-neutral, and radical reactions. Results are summarized in Table~\ref{tab:ener}. 
In Figure~\ref{fig:react-prod} we show the structures of reactants and products considered here (we omitted simple products like H$\cdot$, H$_2$, H$_2$O, and CH$_4$).

First, let us consider ion-molecule reactions. The most simple  ones would be 

\begin{eqnarray}
\mbox{NH}_3 + \mbox{CONH}_2^+ \rightarrow \mbox{UreaH}^+ \label{react:NH3CONH2} \\
\mbox{NH}_4^+ + \mbox{HNCO} \rightarrow \mbox{UreaH}^+, \label{react:NH4CONH} 
\end{eqnarray}

which correspond to the condensation of
NH$_3$ with CONH$_2^+$ or NH$_4^+$ with HNCO, both being exothermic. However, the urea eventually formed needs an environment to keep the excess energy or, alternatively, can be stabilized by
spontaneous emission of a photon (corresponding to a radiative association reaction).
In the gas phase, there is no environment to take the energy of the formed protonated urea, so there are two competing routes: (i) unimolecular dissociation and (ii) radiative decay.
The rate constants for the unimolecular dissociation of protonated urea were previously reported by~\cite{Spezia2009} and they are, for energies in 40--300~kcal/mol range, in the ps-ns timescale. Spontaneous
emission rate constants are an order of magnitude larger (between 40 and 650~s$^{-1}$ in the same energy range), 
suggesting that it is unlikely that the urea would be stabilized in this way. We should note  that this same previous work reported that 
the NH$_3$ + CONH$_2^+$ reaction channel would be barrierless (the original study focused on the unimolecular decomposition of protonated urea).
 
We now consider ion-molecule reactions, which produce two products, such that the final relative translational energy can take the excess energy and stabilize the products.
First, in analogy with reactions proposed for the gas-phase bi-molecular synthesis of formamide (\cite{Redondo2014}) and glycine (\cite{Barrientos2012}), 
we have considered  the following reactions
\begin{eqnarray}
\mbox{NH}_3\mbox{OH}^+ + \mbox{HCONH}_2 \rightarrow \mbox{UreaH}^+ + \mbox{H}_2\mbox{O} \label{react:NH3OHform}\\
\mbox{NH}_2\mbox{OH}_2^+ + \mbox{HCONH}_2 \rightarrow \mbox{UreaH}^+ + \mbox{H}_2\mbox{O} \label{react:NH2OH2form} \\
\mbox{NH}_2\mbox{OH} + \mbox{HCOHNH}_2^+ \rightarrow \mbox{UreaH}^+ + \mbox{H}_2\mbox{O} \label{react:NH2OHformH}
\end{eqnarray}

All of them are strongly exothermic and exergonic, so we better investigated the corresponding PES to obtain information on the barrier energies. These PESs are reported in section~\ref{sec:mech}.

\begin{table*} [hptb]
\caption[]{Energetics: $\Delta$E with ZPE correction and $\Delta$G calculated at 15~K as obtained from MP2/aug-cc-pVTZ//CCSD(T)/aug-cc-pVTZ calculations. Activation energies, E$^a$ (with
ZPE correction), as obtained at the same level of theory for the different path investigated are shown. When a pathway was not characterized (see text) it is marked with --.
For UreaH$^+$ we report the energy formation for the most stable isomer.}
 \label{tab:ener}
 \centering
\begin{tabular}{l l c c c}
            \hline \hline
            &   Reaction                                                                                                &  $\Delta$E(kcal/mol)                    & $\Delta$G (kcal/mol)  & E$^a$ (kcal/mol) \\
            \hline
            
            \hline
  &          Ion-molecule                                                                                               &                                               &                                               &        \\
            \hline
 (\ref{react:NH3CONH2})         & NH$_3$   + CONH$_2^+$ $\rightarrow$  UreaH$^+$                                                 &     -50.48                             &-50.30                                         &   (-7.82)$^a$ \\
 (\ref{react:NH4CONH})  & NH$_4^+$ + HNCO $\rightarrow$  UreaH$^+$                                                              &      -20.31                                     &-20.14                 &   (23.15)$^a$ \\
 (\ref{react:NH3OHform})        & NH$_3$OH$^+$ + HCONH$_2$ $\rightarrow$ UreaH$^+$ + H$_2$O                      &       -71.60                          & -70.10                                  &  17.60 \\
 (\ref{react:NH2OH2form}) & NH$_2$OH$_2^+$ + HCONH$_2$ $\rightarrow$ UreaH$^+$ + H$_2$O                        &      -97.40                                   & -97.30                          &  -8.04 \\
 (\ref{react:NH2OHformH}) & NH$_2$OH + HCOHNH$_2^+$ $\rightarrow$ UreaH$^+$ + H$_2$O                        &     -70.49                            &-70.40                                 &   18.80 \\
 (\ref{react:NH4HCONH2}) & NH$_4^+$ + HCONH$_2$ $\rightarrow$ UreaH$^+$ + H$_2$                                   &      -4.89                                    &-4.73                  &   85.18 \\
 (\ref{react:NH3HCOHNH2}) & NH$_3$ + HCOHNH$_2^+$ $\rightarrow$ UreaH$^+$ + H$_2$                                 &     -12.41                                    &-12.26                 &  77.66 \\
 (\ref{react:CH3NH2FormH}) & CH$_3$NH$_2$ + HCOHNH$_2^+$ $\rightarrow$ UreaH$^+$ + CH$_4$                 &      -35.47                               &     -35.44                                &  70.58 \\
 (\ref{react:CH3NH3Form}) & CH$_3$NH$_3^+$ + HCONH$_2$ $\rightarrow$ UreaH$^+$ + CH$_4$                     &    -16.94                                &       -16.89                          & 89. 13\\     
\hline
 & Neutral-neutral                                                                                                              &                                               &                                       &       \\
\hline
 (\ref{react:neutralNH3CONH}) & NH$_3$ + HNCO $\rightarrow$ Urea                                                                        &    -14.78                                       &-14.64                                 & -- \\
 (\ref{react:neutralNH2OHHCOH2}) & NH$_2$OH + HCONH$_2$ $\rightarrow$ Urea + H$_2$O                                        &   -57.46                              &-57.38                                         & 60.26 \\
 (\ref{react:neutralNH3HCONH2}) & NH$_3$ + HCONH$_2$ $\rightarrow$ Urea + H$_2$                                                   &    0.61                                       &0.75                           & --\\
 (\ref{react:neutraCH4}) & CH$_3$NH$_2$ + HCONH$_2$ $\rightarrow$ Urea + CH$_4$                                  &  -22.44                                       & -22.79                          & -- \\
\hline
 & Radical                                                                                                                      &                                               & \\
\hline                    
 (\ref{react:radicalNH2}) & NH$_2\cdot$ + HCONH$_2$ $\rightarrow$ Urea + H$\cdot$                                        & -1.84                                         &-1.65                          & 16.22 \\
 (\ref{react:radicalNH3}) & NH$_3$ + HCONH$\cdot$ $\rightarrow$ Urea + H$\cdot$                                 &        -11.76                                           &-11.65                 & 6.18 \\ 
            \hline
         \end{tabular}
\\         $^a$ values as from \cite{Spezia2009}.
   \end{table*}
   
We have also considered reactions between NH$_3$, HCONH$_2,$ and CH$_3$NH$_2$, considering in each case a neutral and a protonated species:    
 
\begin{eqnarray}
\mbox{NH}_4^+ + \mbox{HCONH}_2 \rightarrow \mbox{UreaH}^+ + \mbox{H}_2 \label{react:NH4HCONH2} \\
\mbox{NH}_3 + \mbox{HCOHNH}_2^+ \rightarrow \mbox{UreaH}^+ + \mbox{H}_2 \label{react:NH3HCOHNH2}\\
\mbox{CH}_3\mbox{NH}_2 + \mbox{HCOHNH}_2^+ \rightarrow \mbox{UreaH}^+ + \mbox{CH}_4 \label{react:CH3NH2FormH} \\
\mbox{CH}_3\mbox{NH}_3^+ + \mbox{HCONH}_2 \rightarrow \mbox{UreaH}^+ + \mbox{CH}_4 \label{react:CH3NH3Form}
\end{eqnarray}

NH$_3$ (ammonia), HCONH$_2$ (formamide), and CH$_3$NH$_2$ (methylamine ) 
were observed in the ISM by~\cite{ammonia},~\cite{formamide}, \cite{CH3NH2,Fourikis1974},  respectively, so they should be considered as possible reactants. 
Protonated ammonia, NH$_4^+$, cannot be observed for symmetry reasons, but recently the observation of the NH$_3$D$^+$ isotope was reported by~\cite{Cernicharo2013},
strengthening the possibilities of the existence of NH$_4^+$ in the ISM.
Also in this case, all the corresponding reactions are exothermic and exergonic, while the energy and free energy differences are smaller than for reactions~\ref{react:NH3OHform} --~\ref{react:NH2OHformH}.
It is worth investigating the PES also for those reactions in order to localize the barrier energies.

We then considered the following neutral-neutral reactions:

\begin{eqnarray}
\mbox{NH}_3 + \mbox{HNCO} \rightarrow \mbox{Urea} \label{react:neutralNH3CONH}\\
\mbox{NH}_2\mbox{OH} + \mbox{HCONH}_2 \rightarrow \mbox{Urea} + \mbox{H}_2\mbox{O} \label{react:neutralNH2OHHCOH2} \\
\mbox{NH}_3 + \mbox{HCONH}_2 \rightarrow \mbox{Urea} + \mbox{H}_2 \label{react:neutralNH3HCONH2} \\
\mbox{CH}_3\mbox{NH}_2 + \mbox{HCONH}_2 \rightarrow \mbox{Urea} + \mbox{CH}_4 \label{react:neutraCH4}
\end{eqnarray}

Reaction~\ref{react:neutralNH3CONH} is exothermic and exergonic, but in the gas phase it  also has  the problem of stabilization by radiative emission. The radiative emission rate constant 
is in the 40--350~s$^{-1}$ range for energies in the 40--300~kcal/mol range, similar to protonated urea values, 
suggesting that it would be easier to dissociate back to reactants than to stabilize the products by emitting infra-red light. We thus do not follow this reaction.
Reaction~\ref{react:neutralNH3HCONH2} is slightly endothermic and endergonic and thus the barrier will also be positive, and thus we did not study the details of the corresponding PES.
Finally, reactions~\ref{react:neutralNH2OHHCOH2} and~\ref{react:neutraCH4} are exothermic and exergonic, even if, as expected, less so than corresponding ion-molecule reactions. 
Generally, barriers for neutral-neutral reactions are higher (and positive) than corresponding ion-molecule reactions, because in the latter the electrostatic barrier corresponding to the approaching of the reactants 
is lowered by ion-dipole attraction, while in neutral-neutral, weaker dipole-dipole interactions  have a negative contribution to the interaction (\cite{levine}). 
We will show the PES of the most exothermic and exergonic reaction~\ref{react:neutralNH2OHHCOH2}  in order to have information on the barrier in Section~\ref{sec:mechneutral}.

Finally, inspired by the recent work by~\cite{Barone2015} on formamide, we studied two simple radical reactions:

\begin{eqnarray}
\mbox{NH}_2\cdot &+& \mbox{HCONH}_2 \rightarrow \mbox{Urea} + \mbox{H}\cdot \label{react:radicalNH2} \\
\mbox{NH}_3 &+& \mbox{HCONH}\cdot \rightarrow \mbox{Urea} + \mbox{H}\cdot  \label{react:radicalNH3}
\end{eqnarray} 

Also in this case the reactions are exothermic and exergonic, while values are smaller than ion-molecule reactions~\ref{react:NH3OHform}--~\ref{react:NH2OHformH}, but also smaller than 
the neutral-neutral reaction~\ref{react:neutralNH2OHHCOH2}. We will study the PES associated to these radical reaction in Section~\ref{sec:radical}.

To conclude, almost all the reactions reported have a favorable thermochemistry, such that it will be worth investigating for most of them the details of the PES in order to locate the corresponding barriers. We should note
that in both ion-molecule and neutral-neutral reactions, the ones in which a water molecule is formed have the most negative energy and free energy, similarly to what has been reported in previous studies of 
formamide (\cite{Redondo2014}) and glycine (\cite{bohme,Barrientos2012}).

\begin{figure*} [hptb]
\begin{center}
\includegraphics[width=1.0\textwidth]{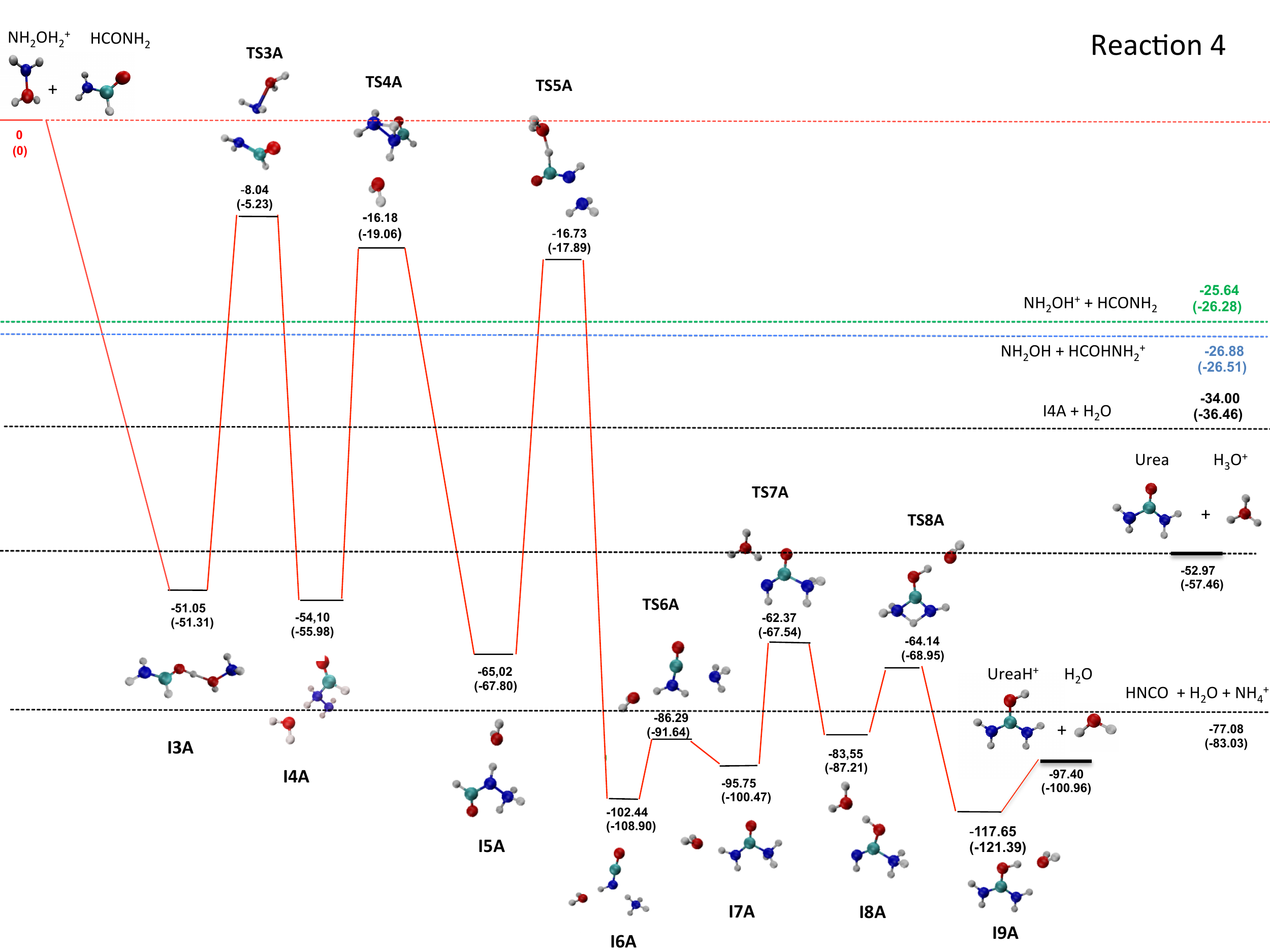}\\
\end{center}\caption[] {Energy profiles, in kcal/mol, for the ion-molecule reaction NH$_2$OH$_2^+$ + HCONH$_2$, 
as obtained by CCSD(T)/aug-cc-pVTZ//MP2-aug-cc-pVTZ  and
MP2/aug-cc-pVTZ (in parenthesis) calculations. These values contain zero-point energies (ZPE). In the final step we report the energy of the cluster, the final product energy being reported in Table~\ref{tab:ener}.}
\label{figure1}
\end{figure*}

\subsection{Mechanisms for ion-molecule reactions}\label{sec:mech}

We now report the PES in terms of connected minima and TSs for the ion-molecule reactions considered previously. First, we start with reactions~\ref{react:NH3OHform}--\ref{react:NH2OHformH}, which are responsible for the formation of protonated urea and neutral water, 
reaction~\ref{react:NH2OH2form} being in particular the most thermodynamically favored among all the
reactions considered in the present work. As we will see, it transpires that this last has a reaction channel leading to the formation of protonated urea (and not only), which is
barrierless. The reaction pathway diagram is shown in Figure~\ref{figure1}.
Investigating in more detail this pathway, we can observe that the interaction between 
NH$_2$OH$_2^+$ with HCONH$_2$ leads to the formation of the complex I3A, which shows a
 hydrogen bond between the carbonyl oxygen of formamide and the hydrogen atom of the amine group of protonated hydroxylamine. 
This complex I3A can then evolve to the I4A intermediate through the transition state TS3A that represents the highest energy barrier (-8.04 kcal/mol with respect to reactants) of this reaction pathway.
 The mechanism that leads to I4A implies the electrophilic attack of the amino group of hydroxylamine to the amino group of formamide. This results in a simultaneous formation of a  N--N bond and in the cleavage of the N--O bond 
 with consequent formation of a water molecule.
 We should note that to form I4A, the system has to pass the TS3A barrier, which is higher in energy than the exit channel leading to NH$_3$OH$^+$ + HCONH$_2$ and NH$_2$OH + 
 HCOHNH$_2^+$ products, shown as
 asymptotic lines in Figure~\ref{figure1}. 
 The system has thus different possibilities and the branching ratio will depend on the dynamical evolution of I3A. If a purely statistical kinetic regime will hold, then the exit channels will be likely favored, but
 a fraction of trajectories would be able to pass the TS3A, forming I4A. Dynamical calculations will be useful in the future to better understand this aspect.

If I4A intermediate is obtained then it can evolve to I5A through hydrogen migration from the secondary to the primary amino group, through TS4A. The barrier of this and the following TSs is lower than the previous one (TS3A). 
The concerted mechanism, responsible for the formation of  I6A through TS5A,  involves the hydrogen abstraction reaction of the C atom by the water molecule and the simultaneous cleavage of the N--N bond. 
This mechanism leads to the formation of I6A, a cluster made by  isocyanic acid, protonated ammonia, and water. 
We should note that all these three species have been detected in the Sgr B2 cloud (\cite{water,isocyanic}); in the case of ammonium, 
the isotope NH$_3$D$^+$ was detected by~\cite{Cernicharo2013}. It is thus possible that this cluster
breaks into its constituents, being a possible barrierless pathway also for the formation of neutral isocyanic acid. In fact, the energy of the three separated species is -77.08~kcal/mol with respect to reactants 
(see the asymptotic line in Figure~\ref{figure1}), 
such that the I6A needs only 25.36~kcal/mol to break into these pieces, which is comparable with the energy needed to pass the TSs and obtain protonated urea + H$_2$O.

From I6A the formation of the more stable isomer of protonated urea takes place through a three-step mechanism. In the first step, the hydrogen abstraction reaction of the ammonium ion by the imino group of the isocyanic acid 
induces the nucleophilic attack of the ammonia to  the isocyanic carbon atom, leading to the formation of N-protonated urea through TS6A.
In the second step, the hydrogen migration, assisted by water which becomes hydronium ion, H$_3$O$^+$, in the transition state TS7A  from the amino-group of N-protonated urea to its oxygen atom, produces I8A.
In the third step, the more stable isomer of protonated urea is produced from I8A through the transition state TS8A, where a hydrogen migration takes place. 

\begin{figure*} [htpb]
\begin{center}
\includegraphics[width=1.0\textwidth]{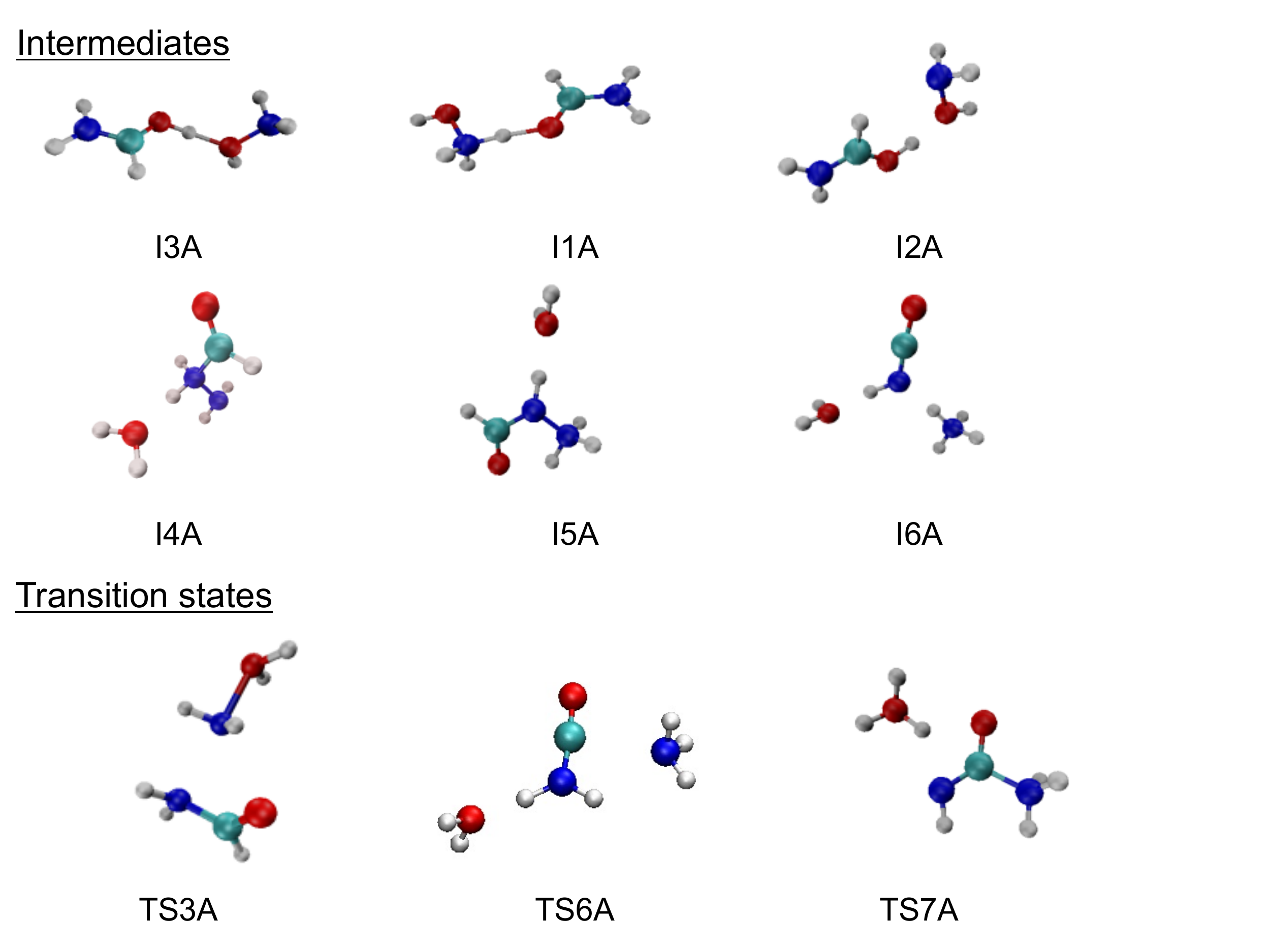}\\
\end{center}\caption{Structures of most relevant intermediates and transition states as obtained from reactions~\ref{react:NH3OHform},~\ref{react:NH2OH2form},  and~\ref{react:NH2OHformH}.
Oxygen atoms are in red, nitrogen in blue, carbon in cyan, and hydrogen in white.}
\label{fig:int_ts}
\end{figure*}

We should note that TS7A is actually a cluster composed of H$_3$O$^+$ and an isomer of neutral urea. The products urea + H$_3$O$^+$ are 52.97~kcal/mol more stable than NH$_2$OH$_2^+$ + HCONH$_2$ 
reactants (see the corresponding asymptotic line in Figure~\ref{figure1}) and the system
has thus enough energy to take a different path instead of the minimum energy path connecting TS7A to I8A (and then to UreaH$^+$ + H$_2$O), thus forming neutral urea. However, when optimizing the neutral
HNCONH$_3$ species of TS7A, we obtained spontaneously isocyanic acid (HNCO) and ammonia. 
 The structures of relevant intermediates and transition states along this pathway are reported in Figure~\ref{fig:int_ts}.

The PESs for reactions~\ref{react:NH3OHform} and~\ref{react:NH2OHformH} show a positive and high activation energy and the corresponding diagrams are reported in Figure~\ref{fig:initialNH3OH}
as Path A and B, respectively. The corresponding barriers of~17.6 and 18.8~kcal/mol for Path A and B, respectively, make the two reactions impossible  under low temperature conditions.
Details are given in Appendix~\ref{app:ionmol}.

The results of our PES analysis show that only the reaction involving the less stable isomer of protonated hydroxylamine, NH$_2$OH$_2^+$, with neutral formamide, HCONH$_2$,
leads to urea through a barrierless pathway. This same pathway can be responsible for the formation of isocyanic acid, which was found as an intermediate of the full pathway. 
Of course, once the I3A complex is formed from a NH$_2$OH$_2^+$ + HCONH$_2$ reaction, it can evolve, forming protonated urea via TS3A or isomerizing into I2A and I1A, 
in this case reacting back and forming
NH$_3$OH$^+$ + HCONH$_2$ or NH$_2$OH + HCOHNH$_2^+$.

\begin{figure} 
\begin{center}
\includegraphics[width=0.45\textwidth]{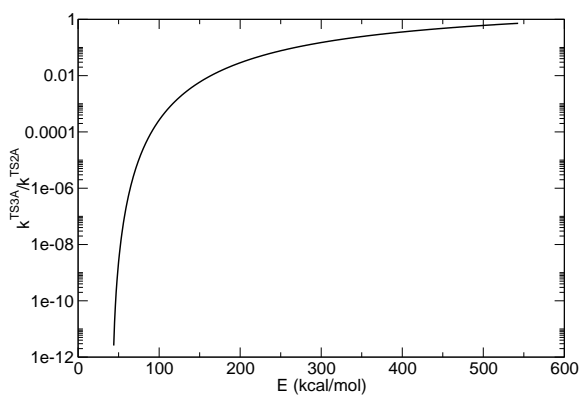}\\
\end{center}\caption{Ratio between k$^{TS3A}$ and k$^{TS2A}$ rate constants corresponding to the activation of I3A complex as function of its internal energy (E).}
\label{fig:kI3A}
\end{figure}
An estimate of the ratio of the two reaction channels can be obtained from the RRKM rate constant to pass the TS3A (k$^{TS3A}$) versus  TS2A (k$^{TS2A}$) transition states, as shown in Figure~\ref{fig:kI3A}. 
The
evolution of the rate constant ratio as a function of internal energy shows that, while for low energy values the pathway initiated by the TS2A isomer is dominant, 
the other pathway becomes increasingly important
as the internal energy of the intermediate increases, with a ratio close to one at higher energies.  
Of course, this is a simple picture that does not take into account the dynamical effect related to the collision between NH$_2$OH$_2^+$ 
and
HCONH$_2$ , and a more detailed study of such reactive scattering events would shed more light on the competition between the different pathways.
In conclusion, both neutral isocyanic acid and protonated urea can be formed from a NH$_2$OH$_2^+$ + HCONH$_2$ reaction under the ISM conditions, even if their formation will be in competition
with the formation of NH$_3$OH$^+$ + HCONH$_2$ and NH$_2$OH + HCOHNH$_2^+$ products.

\begin{figure*} 
\begin{center}
\includegraphics[width=1.0\textwidth]{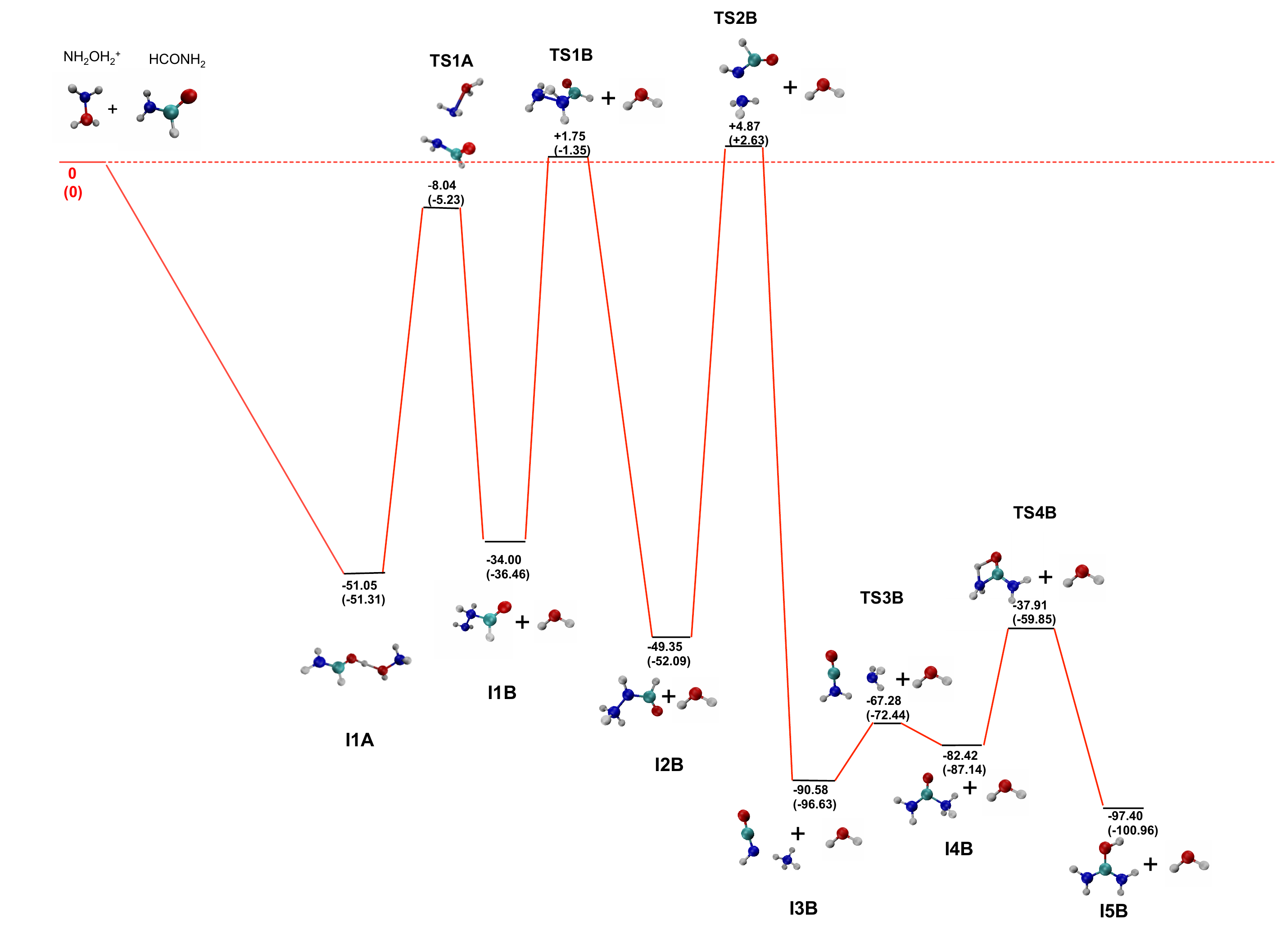}\\
\end{center}\caption{Energy profile, in kcal/mol, for ion-molecule reaction NH$_2$OH$_2^+$ + HCONH$_2$ in which water molecule was separated from ion once formed, 
as obtained by CCSD(T)/aug-cc-pVTZ//MP2-aug-cc-pVTZ  and
MP2/aug-cc-pVTZ (in parenthesis) calculations. These values contain zero-point energies (ZPE).}
\label{figure2}
\end{figure*}

Finally, we investigated the key role of the water molecule in the previously discussed pathway. To this end, we calculated minima and TSs removing the H$_2$O (i.e., calculating separately the energies of  the ions 
and the neutral water molecule) following its formation from intermediate I2B (before the mechanism is by definition the same) up to the final products. Results are shown in Figure~\ref{figure2}.
In the absence of the water molecule, the energy of minima and (more importantly) TSs increases, such that the reaction has a net activation barrier of~4.87~kcal/mol. 
The stability of ion-water clusters during the reaction path is thus crucial, the water molecule acting somehow as a catalyzer in the reaction.

\noindent
We now consider reactions~\ref{react:NH4HCONH2} and~\ref{react:NH3HCOHNH2}, 
involving formamide and ammonia (neutral and protonated). The PES leading to the formation of protonated urea and H$_2$ is shown in Figure~\ref{figure7}. 
The two paths identified have a positive barrier of 85.18 and 77.66~kcal/mol, respectively, and thus they can be disregarded as possible pathways for the formation of urea in the ISM. Details are given in Appendix~\ref{app:ionmol}.

We finally move to ion-molecule reactions~\ref{react:CH3NH2FormH} and~\ref{react:CH3NH3Form} involving methylamine and formamide. 
These reactions are also potentially very interesting since the neutral species of both reactants and products 
were all detected in the ISM (\cite{CH3NH2,urea,formamide,metano}). 
In Figure~\ref{figure6} we report the PES of the two pathways, which have positive barriers of~70.58 and 89.13~kcal/mol, respectively. 
Thus, despite the high exothermicity of  reaction~\ref{react:CH3NH2FormH} ($\Delta$E=-35.47 kcal/mol), it is not likely  to occur in the ISM due to the high activation energy barrier. We discuss details
of these pathways in Appendix~\ref{app:ionmol}.

To summarize, the only ion-molecule reaction that shows a barrierless pathway for the synthesis of protonated urea is reaction~\ref{react:NH2OH2form}, which has as reactant the high energy protonated 
hydroxyl amine (NH$_2$OH$_2^+$).
The same reaction can also lead to neutral isocyanic acid.

\subsection{Mechanisms for neutral-neutral reactions}\label{sec:mechneutral}

We now consider the PES connecting reactants and products for neutral-neutral reaction~\ref{react:neutralNH2OHHCOH2}, which is the most exothermic and exergonic reaction. 
In this case, we found two different pathways, shown in Figure~\ref{figure3}, whose energetic is very similar, both with a positive barrier (61.59 and 60.26~kcal/mol for Path A and B, respectively).
These barriers make the reaction unlikely under ISM conditions and details are thus discussed in Appendix~\ref{app:neutral}.

\subsection{Mechanisms for radical reactions}\label{sec:radical}

Finally, we consider the PES of the two radical reactions~\ref{react:radicalNH2} and~\ref{react:radicalNH3}, 
which are shown in Figure~\ref{figure5} (Path A and B, respectively).
Both reactions occur on a doublet potential energy surface, which is conserved through the reaction. 
The resulting activation energies are of 16.22 and 6.18~kcal/mol, being too high for thermal reactivity in the ISM. 
Details of the PES of these reactions are thus reported in Appendix~\ref{app:radical}.

\section{Conclusions and outlooks}\label{sec:concl}

In this work we have investigated, by highly correlated quantum chemistry calculations, the thermodynamic and kinetic possibility of forming urea in ISM conditions through ion-molecule, neutral-neutral, and radical reactions.
Recently a tentative detection of urea in the ISM was reported by~\cite{urea} and this is the first quantum chemistry study devoted to understanding possible gas-phase reactions concerning the formation of this molecule, 
while a recent theoretical chemistry study relevant to the astrophysical presence of urea considered the stability of different urea isomers, concluding 
that the  (NH$_2$)$_2$CO form is the one likely to be observed in the ISM (\cite{fourre2016}). 
 
We have summarized in Table~\ref{tab:ener} the activation energies, E$^a$, of the different pathways chosen on the basis of our calculations -- here we consider as E$^a$ the energy difference between the reactants and
the  transition state highest in energy along the given pathway. 
Only  the reaction between NH$_2$OH$_2^+$ (the high energy tautomer of protonated hydroxylamine) and formamide is possible through a barrierless mechanism, suggesting the possibility of this happening
in the ISM. The NH$_2$OH$_2^+$ reactant, thus, seems to be a key species, since it was also suggested to be a reactant for the formation of formamide in a recent ion-molecule collisional study (\cite{Spezia2016}). 
In line with what has been suggested
by~\cite{Pulliam2012}, who reported the non-observation of neutral hydroxylamine, it is possible that the protonated species is present in the ISM. 

Protonated urea can give neutral urea via a typical dissociative recombination reaction:

\begin{equation}
\mbox{UreaH}^+ + \mbox{e}^- \rightarrow \mbox{Urea} + \mbox{H}\cdot \label{react:dissrecomb}
\end{equation}

Of course, the addition of one electron to protonated urea can lead also to different fragmentation products of the neutral radical species formed. 
This structure, however, is a (local) minimum, as reported by MP2 geometry optimization, and
to further react, losing H$\cdot$ or other species,  it will need energy and thus it will depend on the internal energy of the molecules.

Another way of forming neutral urea would be by proton exchange with a species, X, with higher proton affinity (PA):

\begin{equation}
\mbox{UreaH}^+ + \mbox{X} \rightarrow \mbox{Urea} + \mbox{HX}^+ \label{react:prot_exchange}
\end{equation}

The  PA of urea was determined experimentally by~\cite{zheng2002} to be 868.4~$\pm$~2.5~kJ/mol, and our calculations report a value of 868.4~kJ/mol at 15~K and 862.5~kJ/mol at 298~K, which is
in good agreement with the experimental determination. This value of PA is relatively high with respect to values known for molecules present in Sgr B2 (data are summarized 
in Table~\ref{tab:PA}), so that only few species, like cyclopropenylidene (PA~=~915.1~kJ/mol), iron oxide (PA~=~907~kJ/mol), methylamine (PA~=~899~kJ/mol), 
ethanimine (PA~=~884.6~kJ/mol), and C$_2$S (PA~=~869.6~kJ/mol) are compatible with a proton transfer reaction from protonated urea.   

Furthermore, we have pointed out that the barrierless mechanism needs  the H$_2$O water molecule,
once formed before the ion isomerizes to the urea structure, to stay close to this ion in order to decrease the proton transfer barriers. 
While the present calculations are done in the gas phase, this result could be a suggestion to
study the role of catalytic ice on these reactions, which could largely favor the reactivity, as pointed out from calculations studying the synthesis of  other species (\cite{Woon99,Woon2001,Woon2002,Koch2008,Chen2011,Rimola2012,Rimola2014}).
Finally, we should note that the same reaction pathway responsible for the formation of protonated urea can also lead to isocyanic acid (here directly the neutral form), which was also observed in the ISM (\cite{isocyanic});
structures of this part of the PES can also lead to H$_3$O$^+$ and neutral urea, while the isomer is not stable and, in our static calculations, spontaneously formed isocyanic acid and ammonia. 

Dynamical-based
calculations would be interesting for future works to better elucidate some possible competing reaction pathways, in particular for the NH$_2$OH$_2^+$ + HCONH$_2$ reaction in which the barrierless formation 
of protonated urea is in competition with simple proton transfers forming NH$_2$OH + HCOHNH$_2^+$ or NH$_3$OH$^+$ + HCONH$_2$.
Other reactions show a too high energy barrier such that they should be disregarded.
In conclusion, quantum chemistry calculations show that it would be possible that protonated urea is formed via an ion-molecule reaction under ISM conditions; then  the neutral form can be finally obtained
 via a typical dissociative recombination reaction or from proton transfer to a molecule with higher proton affinity.
 
\begin{acknowledgements}
R.S. and Y.J. thank ANR DynBioReact (Grant No. ANR-14-CE06-0029-01) for support and CNRS program INFINITI (project ASTROCOL) for partial support. R.S. thanks Universidad de Valladolid for an invited professor fellowship.
\end{acknowledgements}

%
\bibliographystyle{aa} 
\bibliography{IOPexp} 
%

%

%

\begin{appendix} 
\section{Complementary reaction diagrams}
Here we report the energy profiles of the reaction pathways, which show activation energy barriers that are too high.
Details on the neutral-neutral reaction pathways are given in the following.

\subsection{Ion-molecule reactions}\label{app:ionmol}
\begin{figure*} [hptb]
\begin{center}
\includegraphics[width=1.0\textwidth]{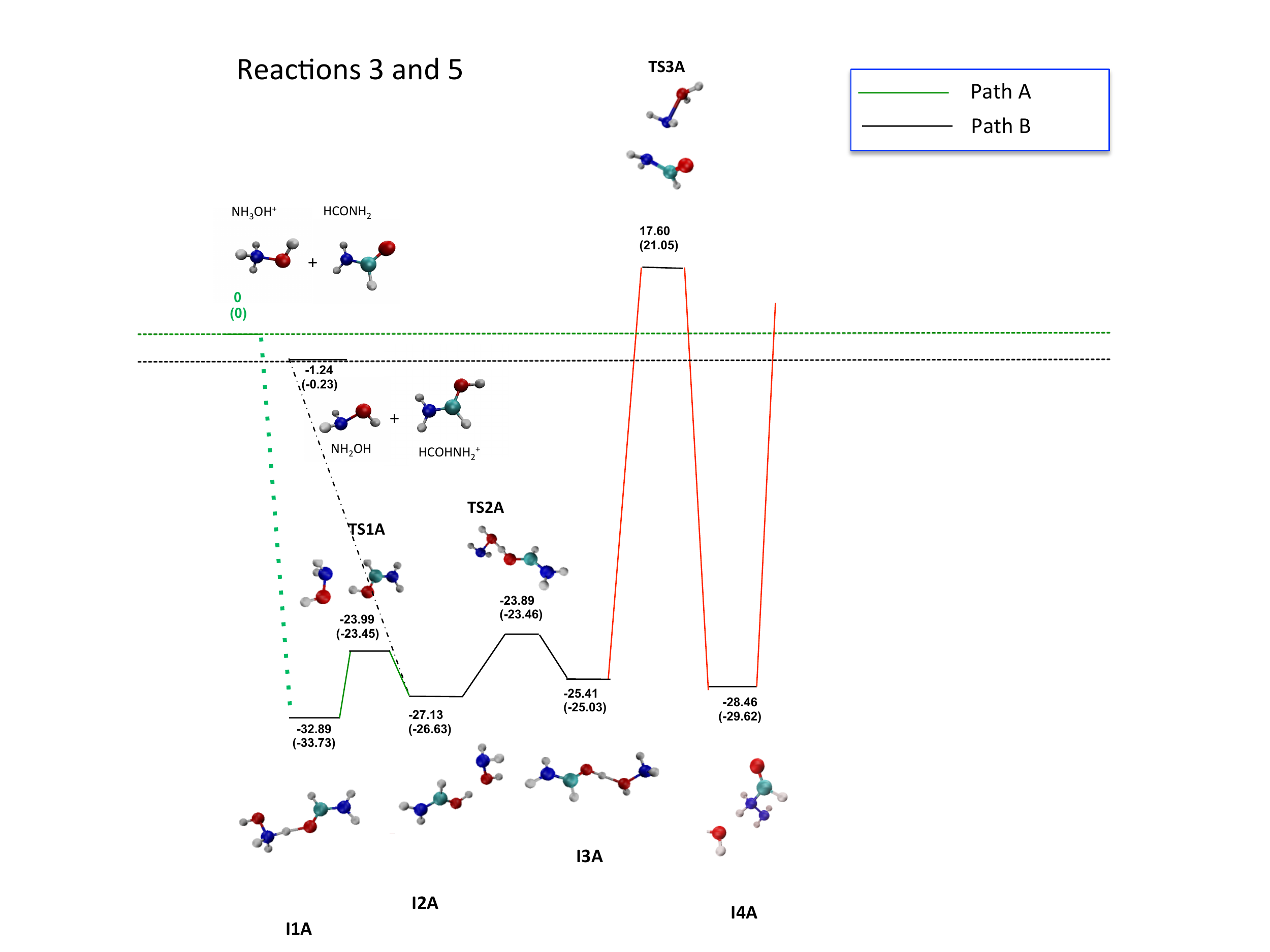}\\
\end{center}\caption[] {Energy profiles, in kcal/mol, for ion-molecule reactions NH$_3^+$OH + HCONH$_2$ (Path A) and NH$_2$OH + HCOHNH$_2^+$ (Path B), 
as obtained by CCSD(T)/aug-cc-pVTZ//MP2-aug-cc-pVTZ  and
MP2/aug-cc-pVTZ (in parenthesis) calculations. These values contain zero-point energies (ZPE). The diagram stops at I4A structure since afterwards it would be the same as Figure~\ref{figure1}.}
\label{fig:initialNH3OH}
\end{figure*}

Here we detail the PES for ion-molecule reactions with energy barriers that are too high, so that these reactions are not doable under low temperature conditions.
First we discuss reactions~\ref{react:NH3OHform} and~\ref{react:NH2OHformH}, reported in Figure~\ref{fig:initialNH3OH} as Path A and B, respectively.
 
The electrostatic interaction between NH$_3$OH$^+$ and the neutral formamide (reaction~\ref{react:NH3OHform}) leads to the formation of the complex I1A. Then, the complex I1A evolves to I2A through a hydrogen migration from the nitrogen to the oxygen atom, involving TS1A. The transition state TS2A connects the two intermediates I2A and I3A, implying the rotation around the torsional angle O-H-O-N. 
The structures of I1A and I2A are also reported in Figure~\ref{fig:int_ts}.
From I3A, the formation of the more stable isomer of protonated urea takes place as described previously. In this case, however, since the reactants are lower in energy, the pathway has a barrier of ~17.6~kcal/mol, 
making this process not possible under low temperature conditions. 

The ion-molecule interaction between neutral hydroxylamine with protonated formamide forms the complex I2A (Path B of Figure~\ref{fig:initialNH3OH}). From I2A, the reaction mechanism for the synthesis of protonated urea is the one already described, with a net activation barrier of~18.8~kcal/mol.

\begin{figure*} [hptb]
\begin{center}
\includegraphics[width=1.0\textwidth]{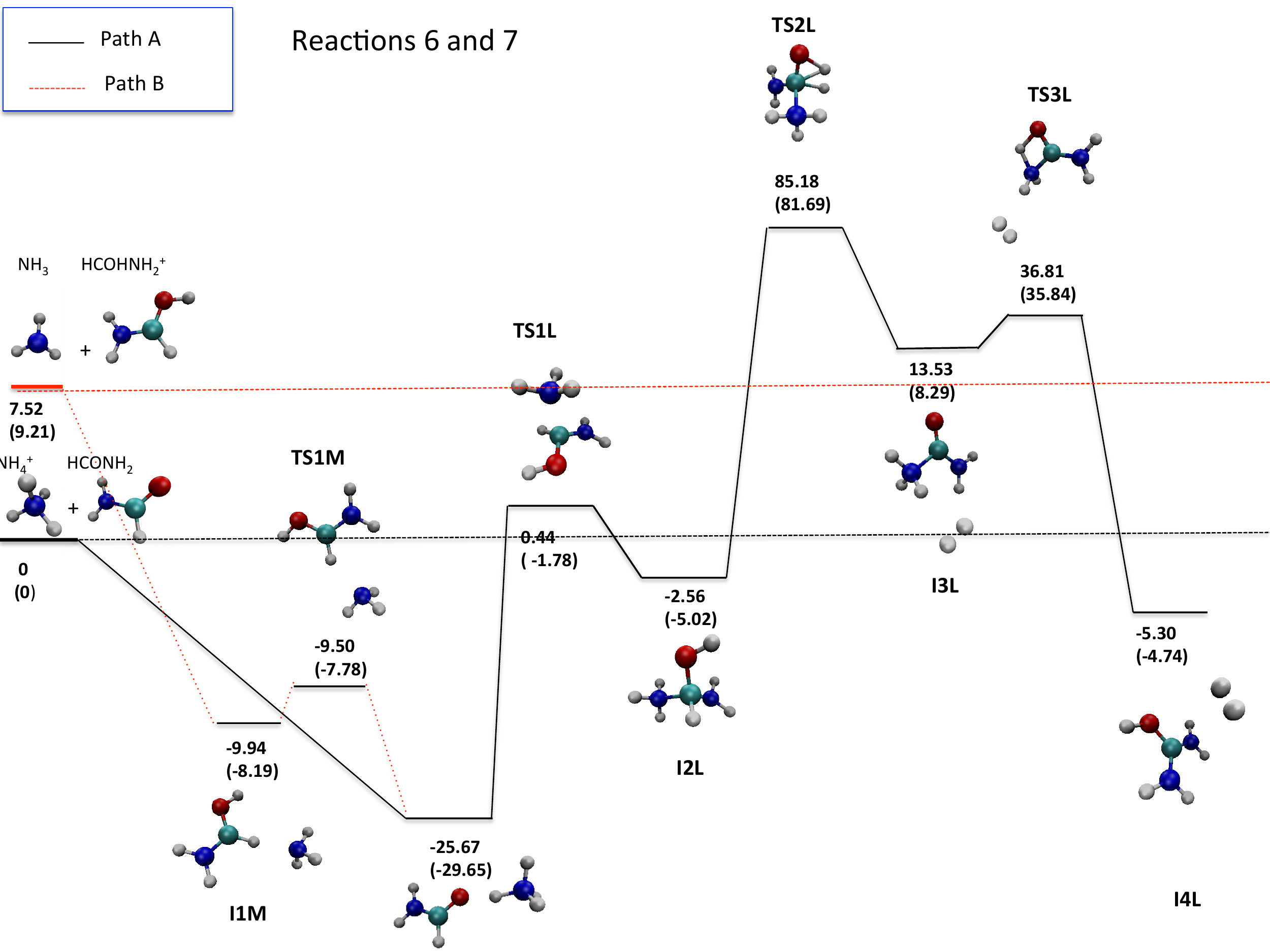}\\
\end{center}\caption[] {Energy profiles, in kcal/mol, for ion-molecule reactions NH$_4^+$ + HCONH$_2$ (Path A) and NH$_3$ +  HCOHNH$_2^+$ (Path B), 
as obtained by CCSD(T)/aug-cc-pVTZ//MP2-aug-cc-pVTZ  and
MP2/aug-cc-pVTZ (in parenthesis) calculations. These values contain zero-point energies (ZPE). In the final step we report the energy of the cluster, the final product energy being reported in Table~\ref{tab:ener}.}
\label{figure7}
\end{figure*}

We now consider reactions~\ref{react:NH4HCONH2} and~\ref{react:NH3HCOHNH2}, 
involving formamide and ammonia (neutral and protonated), whose PES  leading to the formation of protonated urea and H$_2,$ is shown in Figure~\ref{figure7}.
As we can notice from Path A (corresponding to NH$_4^+$ + HCONH$_2$ reactants), the electrostatic interaction between formamide and ammonium ion results in the formation of the complex I1L, where a 
hydrogen bond interaction between the carbonyl oxygen of formamide and ammonium ion takes place.   
The reaction proceeds through TS1L, where the deprotonation of the ammonium ion induces a nucleophilic attack on the carbon atom by the nitrogen atom of ammonia, leading to the formation of I2L.   
From the tetrahedral intermediate I2L, the reaction evolves to I3L by removing  molecular hydrogen through TS2L. This is the rate determinant step of this reaction since it corresponds to the high energy barrier.
Then a proton transfer leads to the final products via  TS3L transition state.
A similar route, in which only the initial steps are different, corresponds to NH$_3$ + HCOHNH$_2^+$ reaction (Path B in Figure~\ref{figure7}). 
In both cases the pathways connecting these reactants to exothermic products
have a highly positive barrier, 85.18 and 77.66~kcal/mol for Path A and B, respectively, and they can be disregarded as possible pathways for the formation of urea in the ISM.

\begin{figure*} [hptb]
\begin{center}
\includegraphics[width=1.0\textwidth]{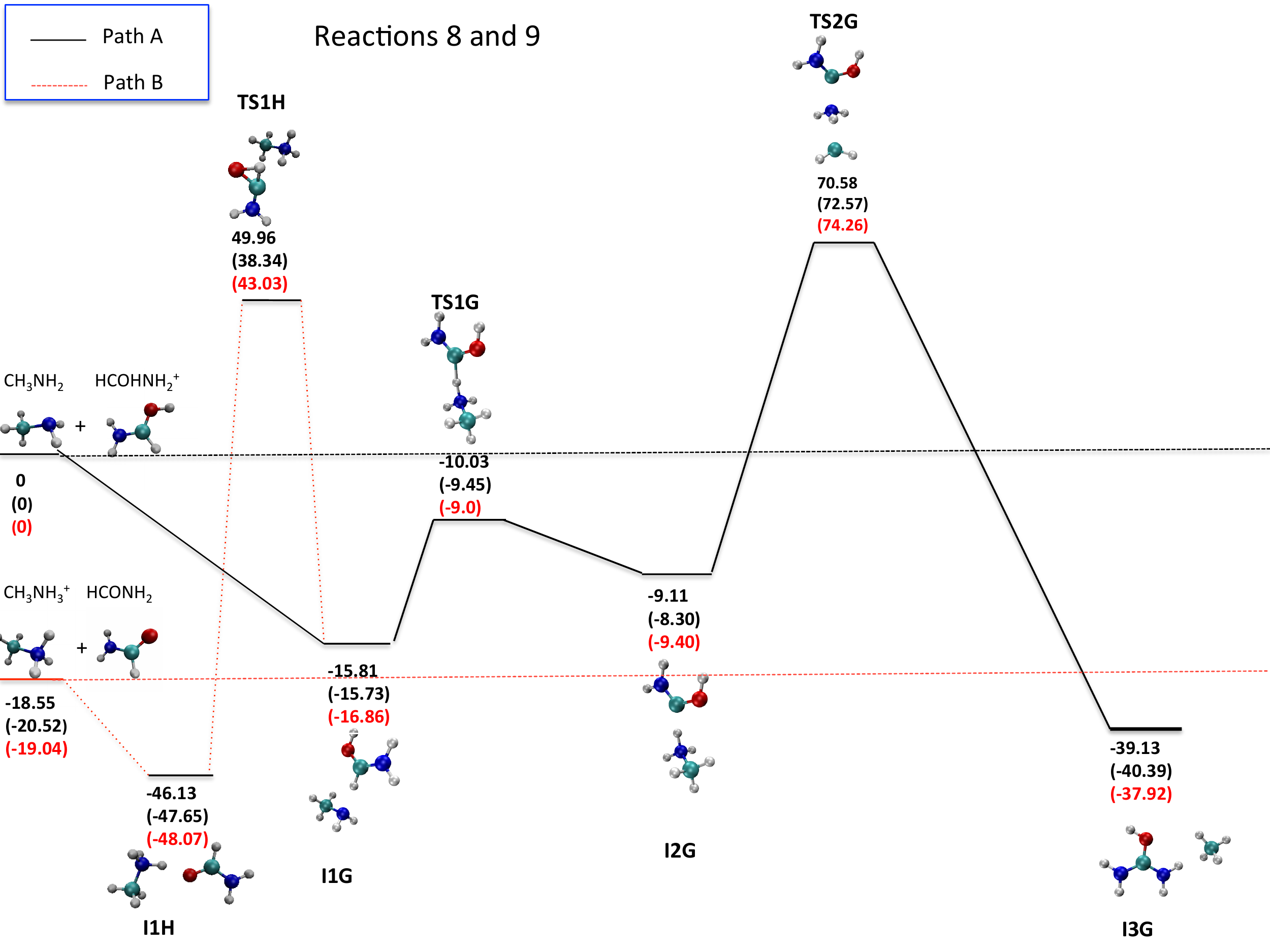}\\
\end{center}\caption[] {Energy profiles, in kcal/mol, for ion-molecule reactions CH$_3$NH$_2$ + HCOHNH$_2^+$ and CH$_3$NH$_3^+$ + HCONH$_2$ as obtained by CCSD(T)/aug-cc-pVTZ//MP2-aug-cc-pVTZ  and
MP2/aug-cc-pVTZ (in parenthesis) calculations. These values contain zero-point energies (ZPE). In red we report values concerning only electronic energies, which show that I2G is lower in energy than the TS1G transition
state for electronic energies, while ZPE inverts the order (see text for more details). In the final step we report the energy of the cluster, the final product energy being reported in Table~\ref{tab:ener}.
}
\label{figure6}
\end{figure*}

Finally, we discuss the PES of
reactions~\ref{react:CH3NH2FormH} and~\ref{react:CH3NH3Form} involving methylamine and formamide, which are reported in Figure~\ref{figure6}.
The reaction between CH$_3$NH$_2$ and HCOHNH$_2^+$ (Path A in Figure~\ref{figure6}) 
is somehow particular, showing a feature that can arise when localizing minima and TSs on the Born-Oppenheimer surface and then adding ZPE correction (as
always in theoretical chemistry). I1G, the complex between the two reactants, and I2G, an intermediate, are connected via a TS, TS1G. Considering only electronic energy (shown in red in Figure~\ref{figure6}),
the TS is higher in energy than the intermediate I2G, while adding the ZPE on reactants, TSs, and products; now the TS is  lower in energy then its corresponding product. We note that the IRC calculation 
connecting,
via the minimum energy path, the TS with the products (in this case I1G and I2G) is done on the Born-Oppenheimer surface, which does not contain ZPE correction. This means that dynamically, even at 0~K, the I1G to I2G
isomerization is barrierless. We will not go more in detail on this aspect since we are interested in the highest energy barrier on each pathway, which in the present case does not correspond to this peculiar TS.
In fact, from I2G the system needs to pass through  TS2G, where the simultaneous cleavage of the C--N bond of methylamine and the formation of a new C--N bond occur.
TS2G is the highest energy TS  of this reaction pathway with an energy barrier of~70.58~kcal/mol.
The other pathway shown in Figure~\ref{figure6} (Path B) differs from Path A by the initial steps. 
First the  complex I1H is formed, then leading to the complex I1G, in which the formamide is now protonated. TS1H is the only structure with a saddle point topology found, since when we try to transfer
a proton from the NH$_3$ group of protonated methylammine to the oxygen atom of formaldehyde, the system always come back to structure I1H. However, the rate determining step is the TS2G, which is in common between the
two pathways.   In fact, once I1G is obtained then  the reaction proceeds as for Path A. 

\subsection{Neutral-neutral reactions}\label{app:neutral}
\begin{figure*} [htpb]
\begin{center}
\includegraphics[width=0.5\textwidth]{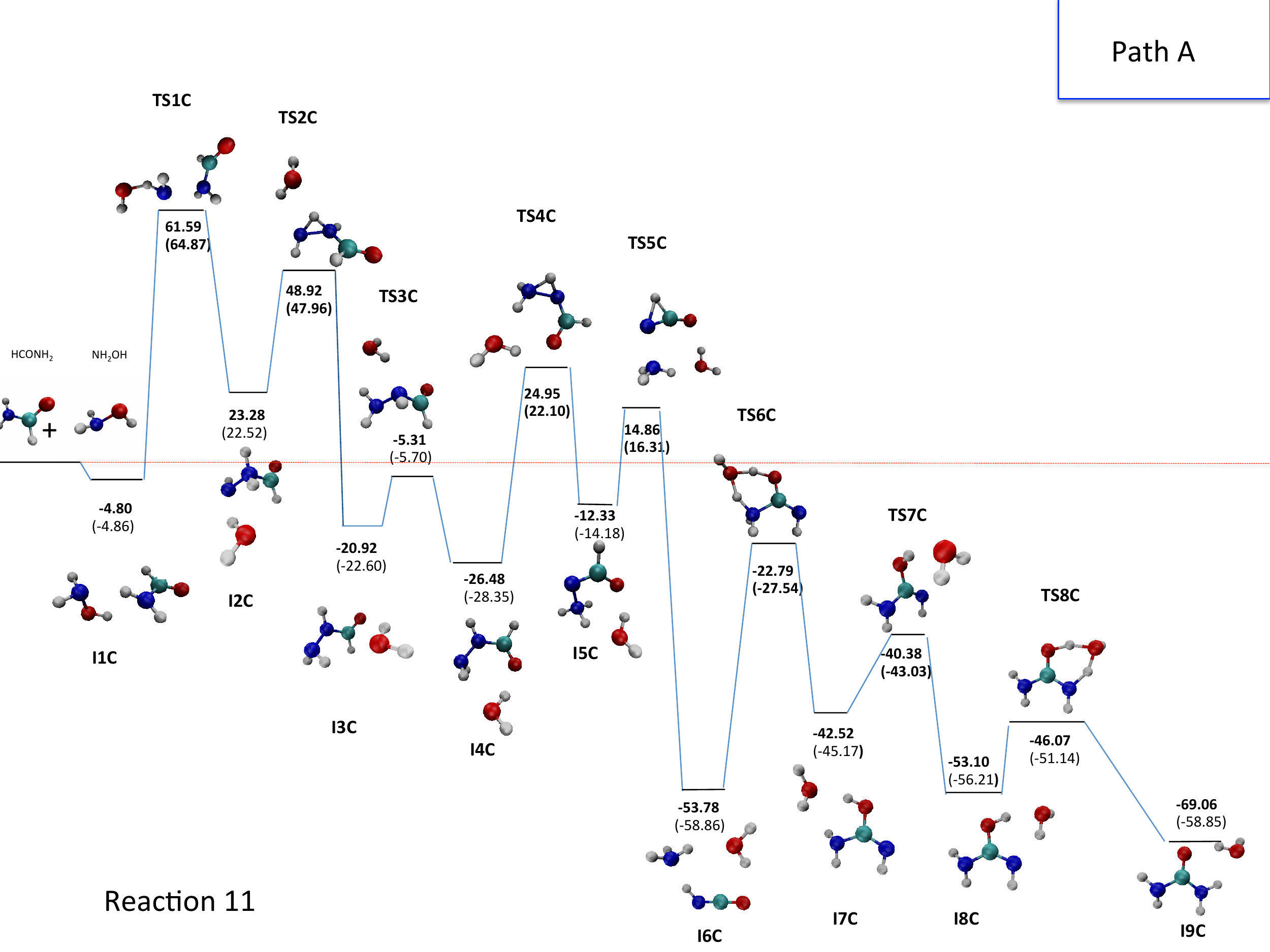}\\
\includegraphics[width=0.5\textwidth]{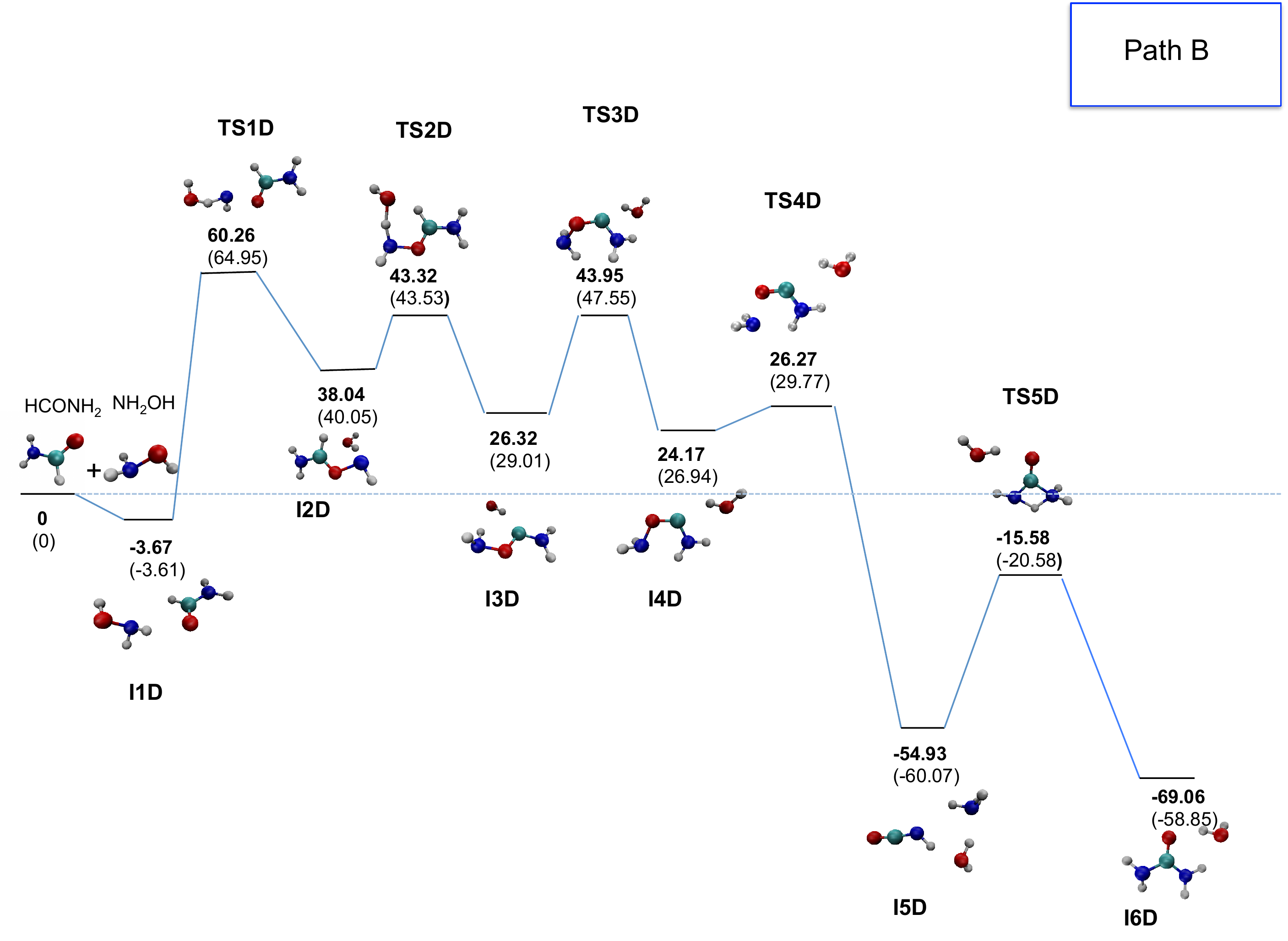}\\
\end{center}\caption{Energy profiles, in kcal/mol, for two pathways of neutral-neutral reaction NH$_2$OH + HCONH$_2$ 
as obtained by CCSD(T)/aug-cc-pVTZ//MP2-aug-cc-pVTZ  and
MP2/aug-cc-pVTZ (in parenthesis) calculations. These values contain zero-point energies (ZPE). In the final step we report the energy of the cluster, the final product energy being reported in Table~\ref{tab:ener}.
Upper panel: Path A; lower panel: Path B. 
}
\label{figure3}
\end{figure*}

In Figure~\ref{figure3} we show the PES corresponding to the neutral-neutral reactions. Two pathways were identified, Path A and Path B, and they both show positive barriers such that they cannot be considered as likely to occur under ISM conditions.
One pathway, Path A in Figure~\ref{figure3},
is the neutral-neutral analog of the one between protonated hydroxylamine and neutral formamide.
The first step of this process is the formation of the complex I1C, resulting in a hydrogen bond interaction between the amine group of the formamide and the amino group of hydroxylamine. 
 The 1,2-hydrogen shift from the nitrogen atom to the oxygen atom in the hydroxylamine and the successive formation of a new  N--O bond leads to the intermediate I2C, through the transition state TS1C. 
 This transition state is 61.59~kcal/mol higher than  the reactants and represents the highest energy barrier  of this reaction path.
The reaction advances through the hydrogen migration from the secondary to the primary amine of I2C and leads to the formation of the intermediate I3C, involving TS2C.
The I3C evolves to the I4C through the TS4C, where a rotation of the aldehyde group around the C--N bond occurs. The I4C progresses through TS5C, where a hydrogen migration from the secondary amino group of I4C to its primary amino group leads to the intermediate I5C.
From I5C the reaction reaches I6C through a concerted  mechanism, via TS5C: a hydrogen atom migration from the carbon atom to the nitrogen atom of the secondary amino group, which induces the N--N bond cleavage 
and the consequent formation of ammonia.
The intermediate I6C is a cluster formed by isocyanic acid, water, and ammonia, similar to what is obtained in the pathway of reaction~\ref{react:NH2OH2form}, here of course all species being neutral. 
From I6C the reaction progresses to I7C through the TS6C, which implies a proton transfer, assisted by water, from the nitrogen atom of the ammonia molecule to the carbonyl oxygen of the isocyanic acid. 
This proton transfer favors the nucleophilic attack by the nitrogen atom on the carbonyl carbon atom.    
Then, the rotation of the alcoholic hydrogen around the C-O bond takes place in TS7C and leads to I8C.
From I8C the urea is produced through TS8C, where a proton transfer, assisted by water, takes place from the oxygen atom to the nitrogen atom.

Path B is shown in Figure~\ref{figure3}.
In this case, the reaction begins with the formation of the complex I1D, which implies a hydrogen bond interaction between the  oxygen of formamide and the hydrogen of the amino group of hydroxylamine, and which is
slightly less stable than the complex I1C of Path A.
 The reaction mechanism, which leads from I1D to I2D, then consists in a 1,2 hydrogen shift from the nitrogen atom to the oxygen atom of the hydroxylamine, with consequent cleavage of the N--O bond.   
The highest barrier for this path is TS1D and it is located~60.26~kcal/mol above the energy of the reactants.
A proton transfer, assisted by water, from the carbon atom to the nitrogen atom leads to the formation of I3D, through TS2D.
Then, I3D evolves to I4D via TS3D, where a rotation of the torsional angle N-O-C-N takes place.
 The hydrogen abstraction reaction of the amino group by the NH$_2$ group through TS4D  leads to the formation of I5D, which is (again)
 a neutral cluster composed by isocyanic acid, ammonia, and water.
 From I5D, the hydrogen migration from the nitrogen atom of ammonia to the nitrogen atom of the amino group induces a nitrogen nucleophilic attack on the carbonyl C atom, leading to I6D, via structure TS5D.

\subsection{Radical reactions}\label{app:radical}
\begin{figure*} [htpb]
\begin{center}
\includegraphics[width=1.0\textwidth]{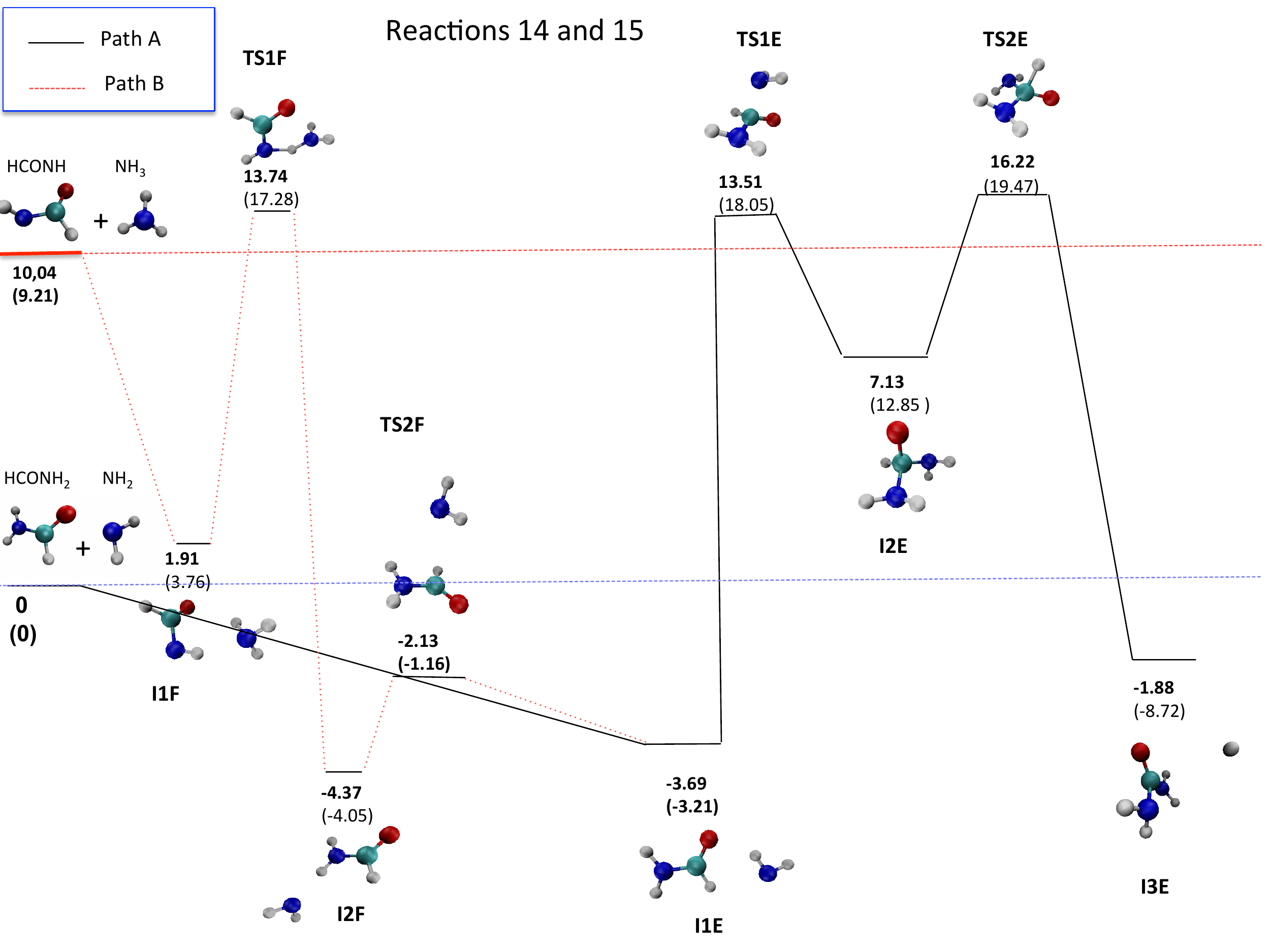}\\
\end{center}\caption{Energy profiles, in kcal/mol, for radical reactions NH$_2$ + HCONH$_2$(black path) and HCONH + NH$_3$(red path) at the CCSD(T)/aug-cc-pVTZ)//MP2/aug-cc-pVTZ and MP2/aug-cc-pVTZ(in parentheses) levels. Zero-point vibrational energy differences are included}
\label{figure5}
\end{figure*}

The PESs corresponding to NH$_2\cdot$ + HCONH$_2$ and NH$_3$ + HCONH$\cdot$ are shown in Figure~\ref{figure5} as Path A and B, respectively. In both cases the activation
energy is positive.
In the case of Path A, the reaction  starts with the formation of the complex I1E, in which the amino radical interacts with the carbonyl group of neutral formamide.
The complex I1E progresses to the tetrahedral intermediate I2E through the TS1E structure, which implies the nucleophilic attack of the nitrogen atom on the carbonyl C atom. The
I2E  evolves to the urea through the TS2E, where the homolytic cleavage of the  C--H bond occurs .
Even if this radical reaction is exothermic (-1.84~kcal/mol), the net activation barrier is~16.22~kcal/mol,  making this process unlikely for ISM conditions.  
The PES profile for Path B, corresponding to the most exothermic and exergonic reaction, is also reported in Figure~\ref{figure5}. At the beginning the  electrostatic interaction between the ammonia and the radical, formamide 
is responsible for the  formation of the complex I1F.
Then, the hydrogen abstraction reaction of the ammonia by the nitrogen atom of radical formamide leads  from I1F to I2F, through TS1F.
 The I2F evolves to I1E through the reorientation of the amino radical around the formamide molecule.
From I1E the system progresses to the products with the same reaction mechanism described previously for Path A.
In this case, the activation barrier of the process is smaller than for Path A but still positive (6.18~kcal/mol), thus making it unlikely for the ISM. 

\section{Proton affinity}
Here we report the proton affinities of molecules observed or for which tentative observations are reported in the Sgr B2 cloud from data listed in the National Institute of Standards and Technology (NIST) web-book. 
For urea  we report also the values
obtained from our calculations at 15 and 298~K, and for hydroxylamine (not present in the experimental data-base) we report calculated values at the same temperatures.
  
\begin{table*}
\caption{Experimental values of available proton affinities (PA) of neutral molecules suggested to be present in the Sgr B2 cloud.}\label{tab:PA}
\centering
\begin{tabular}{llll}
\hline \hline
Molecule & PA (kJ/mol) & Source         & Reference\\
\hline
Urea                    & 868.4~$\pm$~2.5                       & NIST                                          & \cite{zheng2002}        \\
                        & 868.432                               & calculations at T~=~15~K             & this work     \\
                        & 862.549                               & calculations at T~=~298~K            & this work     \\
\hline                                  
Cyclopropenylidene      & 951.1                 & NIST                                          & \cite{hunter98} \\
Iron oxide                      & 907                   & NIST                                          & \cite{hunter98} \\
Methylamine             & 899.0                 & NIST                                          & \cite{hunter98} \\
Ethanimine              & 885.1                 & NIST                                          & \cite{hunter98} \\
C$_2$S                  & 869.6                 & NIST                                          & \cite{hunter98} \\
\hline
Acetamide       & 863.6                         & NIST                                          & \cite{hunter98} \\
N-methylformamide       & 851.3                 & NIST                                          & \cite{hunter98} \\
Methyl isocyanide       & 839.1                 & NIST                                          & \cite{hunter98} \\
Ketene                  & 825.3                         & NIST                                          & \cite{hunter98} \\
Aminoacetonitrile       & 824.9                 & NIST                                          & \cite{hunter98} \\
Formamide       & 822.2                         & NIST                                          & \cite{hunter98} \\
Methyl acetate          & 821.6                 & NIST                                          & \cite{hunter98} \\
Phenol                  & 817.3                 & NIST                                          & \cite{hunter98} \\
Ethylene glycol         & 815.9                 & NIST                                          & \cite{hunter98} \\
Acetone                 & 812                   & NIST                                          & \cite{hunter98} \\
Cyanamide               & 805.6                 & NIST                                          & \cite{hunter98} \\
Isopropyl cyanide       & 803.6                 & NIST                                          & \cite{hunter98} \\
Propylene oxide & 803.3                 & NIST                                          & \cite{hunter98} \\
Ethyl formate           & 799.4                 & NIST                                          & \cite{hunter98} \\
Propyl cyanide          & 798.4                 & NIST                                          & \cite{hunter98} \\
Propenal                        & 797                   & NIST                                          & \cite{hunter98} \\
Ethylcyanide            & 794.1                         & NIST                                          & \cite{hunter98} \\
Phosphorus mononitride & 789.4          & NIST                                          & \cite{hunter98} \\
Propanal                        & 786                   & NIST                                          & \cite{hunter98} \\
Vinylcyanide            & 784.7                         & NIST                                          & \cite{hunter98} \\
Acetic acid             & 783.7                 & NIST                                          & \cite{hunter98} \\
Methyl formate          & 782.5                 & NIST                                          & \cite{hunter98} \\
Methyl cyanide          & 779.2                 & NIST                                          & \cite{hunter98} \\
Silicon monoxide        & 777.8                 & NIST                                          & \cite{hunter98} \\
Ethanol                 & 776.4                 & NIST                                          & \cite{hunter98} \\
Methyl mercaptan        & 773.4                         & NIST                                          & \cite{hunter98} \\
Amino radical           & 773.4                 & NIST                                          & \cite{hunter98} \\
Hydrogen isocyanide      & 772.3                        & NIST                                          & \cite{hunter98} \\ 
Methyl isocyanate       & 764.4                 & NIST                                          & \cite{hunter98} \\
Acetaldehyde    & 768.5                         & NIST                                          & \cite{hunter98} \\ 
Thioformaldehyde        & 759.7                         & NIST                                          & \cite{hunter98} \\
Methanol                & 754.3                         & NIST                                          & \cite{hunter98} \\
Isocyanic acid          & 753                   & NIST                                          & \cite{hunter98} \\
Cyanoacetylene  & 751.2                 & NIST                                          & \cite{hunter98} \\
Methyl-acetylene        & 748                   & NIST                                          & \cite{hunter98} \\
Formic acid             & 742.0                 & NIST                                          & \cite{hunter98} \\ 
Water           & 691                           & NIST                                          & \cite{hunter98} \\
Carbonyl sulfide        & 628.5                 & NIST                                          & \cite{hunter98} \\
N$_2$O                  & 575.2                 & NIST                                          & \cite{hunter98} \\
Nitric oxyde            & 531.8                 & NIST                                          & \cite{hunter98} \\
Hydrogen fluoride       & 484                   & NIST                                          & \cite{hunter98} \\
\hline
\end{tabular}
\end{table*}
%
\end{appendix}
\end{document}